\documentclass{svmult}

\usepackage{makeidx}         
\usepackage{graphicx}        
\usepackage{multicol}        
\usepackage[bottom]{footmisc}

\makeindex             

\begin{document}

\title*{The Apparent Madness of Crowds:\\
{\Large Irrational collective behavior emerging from interactions among rational agents
}}
\titlerunning{The Apparent Madness of Crowds} 
\author{Sitabhra Sinha}
\institute{The Institute of Mathematical Sciences, C. I. T. Campus,
Taramani,\\ 
Chennai - 600 113, India.\\
\texttt{sitabhra@imsc.res.in}}

\maketitle
\noindent
Standard economic theory assumes that agents in markets behave rationally.
However, the observation of extremely large fluctuations in the price
of financial assets that are not correlated to changes in their fundamental
value, 
as well as the extreme instance of financial bubbles and crashes,
imply that markets (at least occasionally) do display irrational
behavior. In this paper, we briefly outline our recent work demonstrating 
that
a market with interacting agents having bounded rationality can
display price fluctuations that are {\em quantitatively} similar to those 
seen in 
real markets.

\vspace{-0.5cm}
\section{Introduction}
\label{sec:1}
\vspace{-0.1cm}
It has long been debated in the economic literature whether markets exhibit 
irrational behavior \cite{albin98}. The historical observations of 
apparent financial ``bubbles", in which the demand (and therefore, the price) 
for certain assets
rises to unreasonably high levels within a very short time
only to come crashing down later \cite{chancellor99}, imply that markets 
act irrationally, because the rapid price changes are not associated with
changes in the fundamental value of the assets. Believers in rational
expectation theory argue that the price rise actually reflects the
market's expectations about the long-term prospect of 
these assets and the large fluctuations are just rapid adjustments
of these expectations in the light of new information \cite{garber90}.
These advocates of the ``efficient market" school of thought claim that 
popular descriptions of speculative mania (e.g., in Ref.~\cite{mackay52})
have been often exaggerated. However, critics point
out that the market's estimate of the long-term value of an asset is
a quantity that cannot be measured, and therefore, it is difficult to verify 
whether historical bubbles were indeed rational outcomes. 

In this paper, we take an intermediate position between these two
opposing camps. We assume that individual agents do behave in a rational
manner, where rationality is identified with actions conducive
to market equilibrium. In other words, rational agents will act 
in such a way that the market is ``balanced'', exhibiting
neither excess demand nor supply. Therefore, we expect only small
fluctuations about the equilibrium when we have a large ensemble
of non-interacting agents. 
In the model presented in this paper, market behavior is described by
the collective decision of many interacting agents,
each of whom choose whether to buy or sell an asset
based on the limited information available to them about its prospects.
In isolation, each agent behaves so as to drive the market to equilibrium.
We investigate the possibility that
interactions between such agents can severely destabilize the market
equilibrium. In fact, we show that when agents are allowed
to modify their interactions with neighbours, based on information about 
their past performance in the market, this results in the market
becoming unbalanced and exhibiting extremely large
fluctuations that are quantitatively similar to those seen in real markets.

\vspace{-0.5cm}
\section{Collective irrationality in an agent-based model}
\label{sec:2}
\vspace{-0.25cm}
In this section, we present an agent-based model of the fluctuation
of demand for a particular asset. The agents are assumed to be operating
under bounded rationality, i.e., they try to choose between buying
and selling the asset based on information about the action of their immediate
neighbors and how successful their previous choices were. The
fundamental value of the asset is assumed to be unchanged throughout
the period. From the ``efficient markets'' hypothesis, we should therefore
expect to see only small departures from the equilibrium. In addition,
the agents are assumed to have limited resources, so that they cannot
continue to buy or sell indefinitely. However, instead of introducing
explicit budget constraints \cite{iori02}, we have implemented 
gradually diminishing 
returns for a decision that is taken repeatedly. 

We assume that all agents are placed on a lattice, each site being occupied
by one agent. An agent can only interact with its immediate neighbors
on the lattice. In the simulations reported here, we have considered
a two-dimensional hexagonal lattice, so that the number of neighbors
is $z = 6$. At any given time $t$, the state of an agent $i$ is fully 
described by two variables: its choice, $S_i^t$, and its belief
about the outcome of the choice, $\theta_i^t$. The choice can be either
{\em buy} ($= + 1$) or {\em sell} ($= - 1$), while the belief can vary 
continuously over a range. The behavior of the agent over time can then
be described by the equations governing the dynamics of $S$ and $\theta$,
\begin{equation}
S_i^{t+1} = {\rm sign} ( {\Sigma}_j J_{ij}^t S_j^t - {\theta}_i^{t}),~
{\theta}_i^{t+1} = {\theta}_i^t + \mu_i S_i^{t+1},
\label{adaptive2}
\end{equation}
where, $J_{ij}^t$ measures the degree of interaction between neighboring
agents. The adaptation rate, $\mu_i$, governs the time-scale of diminishing
returns, over which the agent switches from one choice to another
in the absence of any interactions between agents. The overall
state of the market at any given time is described by the fractional 
excess demand, $M^t = (1/N) \Sigma_j S_j^t$. 

In previous work \cite{sinha04}, we have shown that, if the interactions
between agents do not change over time (i.e., $J_{ij}^t = J,$ a constant),
then $M$ shows only small fluctuations about 0. This accords with the
``efficient markets'' hypothesis that any transient imbalance in the demand
or supply of the asset is quickly corrected through the appropriate 
response of agents, so that the market 
remains more or less in equilibrium.
However, if the agents have access to global information about the
market (i.e., $M$), under certain conditions
this can lead to large deviations from the market
equilibrium. We have previously shown that if $M$ is allowed to affect the
belief dynamics of agents, then the market spends most of the time in states
corresponding to excess demand or excess supply. This kind of two-phase
behavior \cite{sinha05} points to the destabilizing effect of 
apparently innocuous information exchanges in the market.

\begin{figure}[tbp]
\centering
\includegraphics[width=0.49\linewidth,clip]{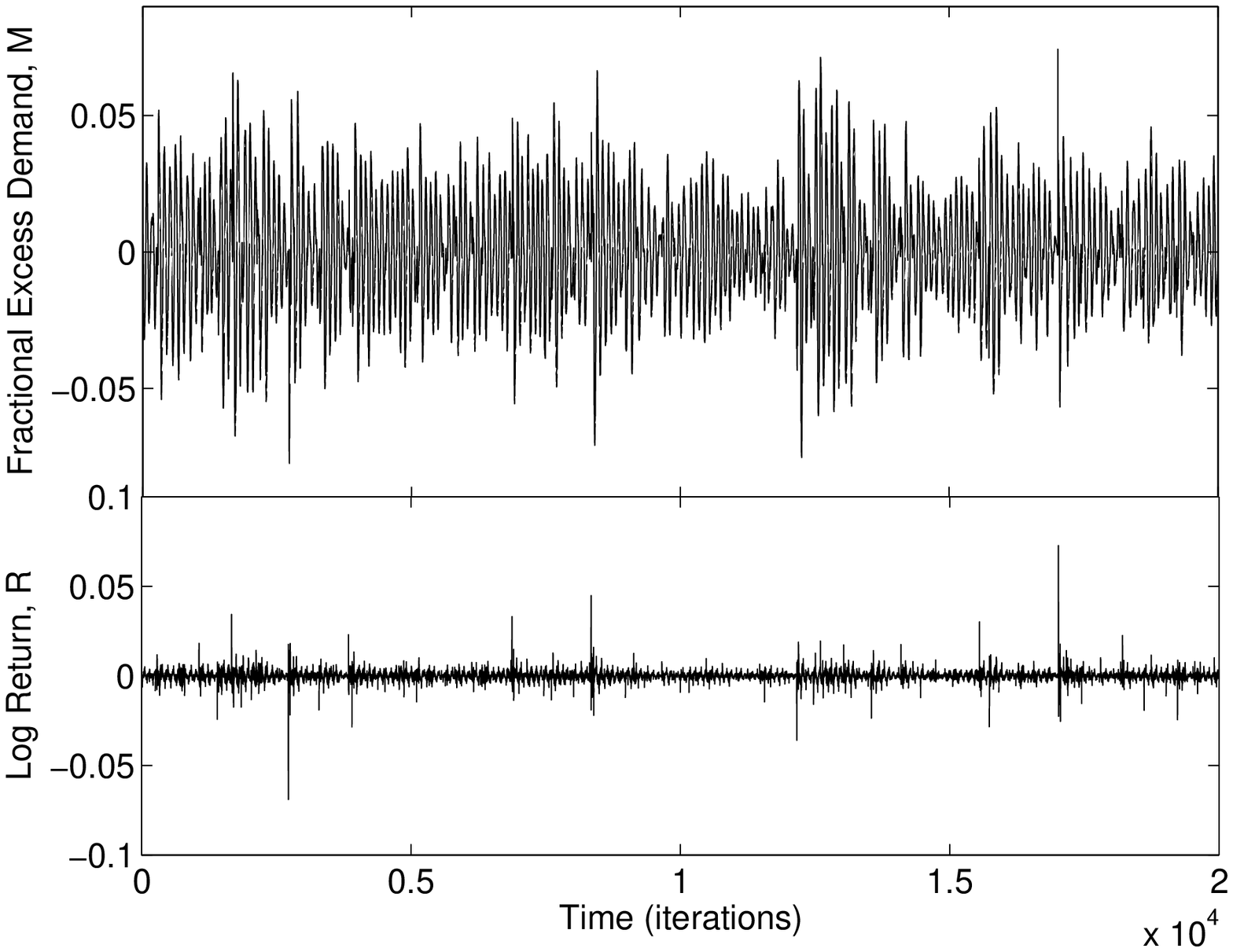}
\includegraphics[width=0.49\linewidth,clip]{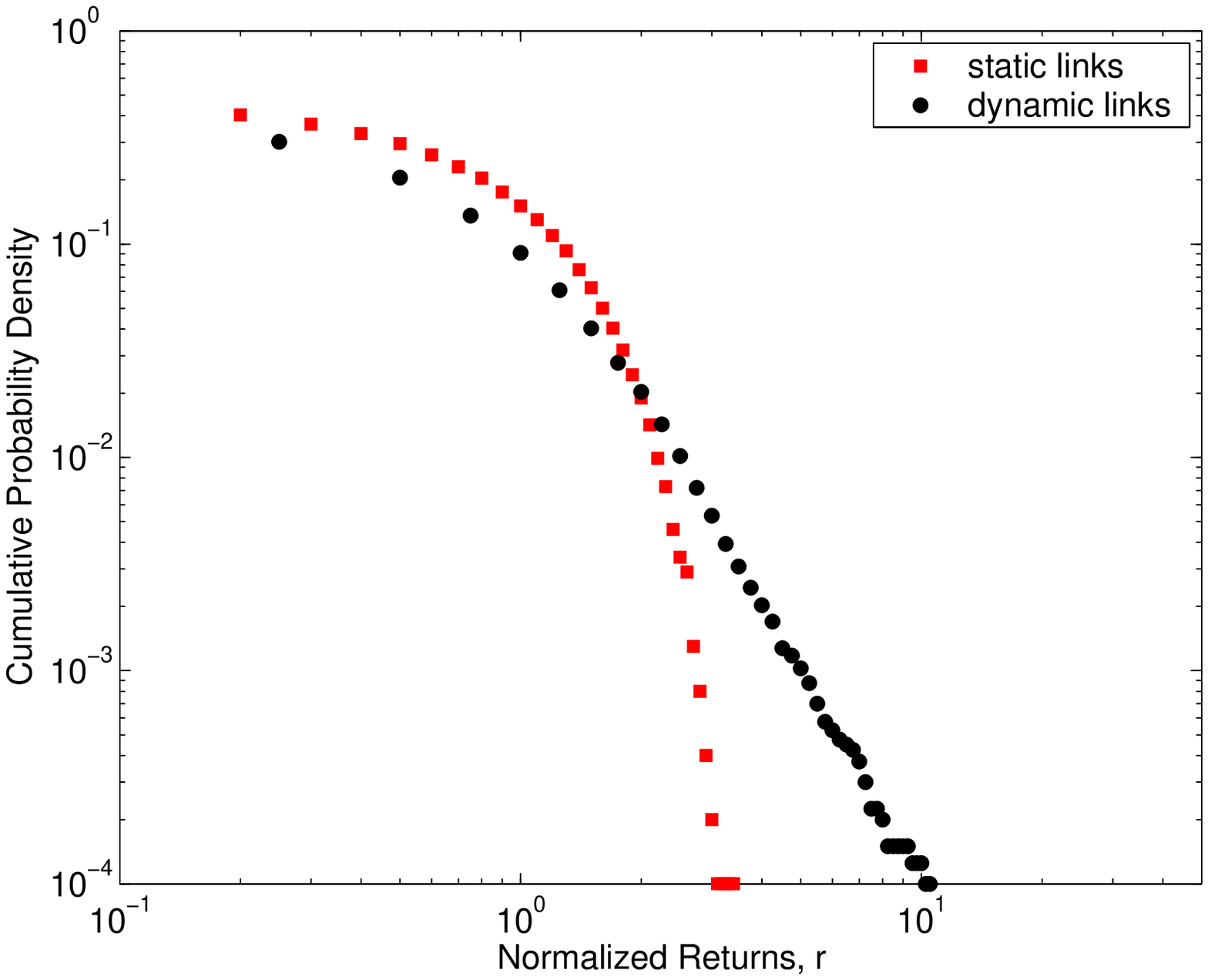}\\
\caption{Spontaneous fluctuations through dynamic interactions in 
a system of $400 \times 400$ agents. The adaptation rates, $\mu_i$, are
chosen from an uniform random distribution between $[0,0.1]$.
(Left) The time series of the fractional excess demand, $M$ (top) and the
corresponding logarithmic return, $R$ (bottom), exhibiting significant
number of large fluctuations. (Right) The cumulative probability density 
function of normalized returns, $r$, comparing the case where the
$J_{ij}$ are fixed in time (squares) with the situation when the
$J_{ij}$ change over time (circles). The latter shows a power-law
decay with an exponent close to 3.}
\label{ssfigure}       
\vspace{-0.15cm}
\end{figure}
Very recently, we have observed that the collective behavior can also be 
destabilized if, instead of affecting the belief dynamics, the 
knowledge of $M$ is used in evolving the structure of interactions $J_{ij}^t$
between neighboring agents. This is implemented by assuming that 
agents seek out the most successful agents in its neighborhood, and 
choose to be influenced by them preferentially. Here, {\em success} is measured
by the fraction of time the agent's decision (to buy or sell) accorded with 
the market behavior. As a rise in excess demand of an asset is taken
to signal its desirability, an agent is considered successful
if it is in possession of an asset that is in high demand.
If an agent $i$ is successful in predicting the market (i.e., its action
in the last round accorded with the majority decision of the collective)
then its interaction structure is unchanged. Otherwise,
its neighboring agents with higher success are identified 
and the link strength between them, $J_{ij}$, is adjusted by an amount
that is proportional to the ratio of
the success of agent $j$ to agent $i$.
This implies that agents with higher success affect the decision process 
of agents with less success, but not the other way around. Finally, 
$J_{ij}$ is normalized such that, for each agent, $\Sigma_i J_{ij} = 1$. 

Fig.~\ref{ssfigure} (left) shows the resulting time 
series of the fractional excess
demand. As the price $P$ of the asset is governed by the demand for it, we can
take $P$ to be linearly related to $M$. This
allows us to quantify the price fluctuations for the asset by calculating 
the logarithmic return of $P$ as $R^t = \ln {P^{t+1}} - \ln {P^{t}}$. 
It is evident that the fluctuations are much larger than what would
have been expected from an uncorrelated random process. This is 
further established when we plot the distribution of the return,
normalized by its standard deviation, and compare it with the
case where the $J_{ij}$ are constant in time (Fig.~\ref{ssfigure}, right). 
While the latter case is
consistent with a Gaussian distribution, the model with adaptive interaction 
dynamics is found to exhibit a return distribution that has a power law tail.
Moreover, the exponent of the cumulative distribution, $\alpha \simeq 3$,
is found to agree {\em quantitatively} with the corresponding values 
observed in actual markets \cite{gopikrishnan98}.
\vspace{-0.5cm}
\section{Conclusions}
\vspace{-0.25cm}
The observation of large price fluctuations (most strikingly during bubbles 
or crashes) 
implies that markets often display instabilities where the demand and
supply are not even approximately balanced.
We have seen in this paper that this is not necessarily inconsistent
with the assumption that individual economic agents are rational. 
A simple agent-based model, where the 
structure of interactions between agents evolve over time based on information 
about the market, exhibits extremely large fluctuations around the market
equilibrium that qualitatively match the fluctuation distribution 
seen in real markets. 

\vspace{-0.2cm}


\printindex
\end{document}